# Ubiquitous Charge Order Correlations in High-Temperature Superconducting Cuprates

Shin-ichi Uchida[1, 2]

[1]Institute of Advanced Industrial Science & Technology, Tsukuba, Japan
[2]Institute of Physics, Chinese Academy of Sciences, Beijing, China

The presence of charge order in high-transition-temperature copper oxides (high-$T_c$ cuprates) was identified a decade ago. Now it is a universally observed order like the antiferromagnetic and the superconducting orders of the cuprates. The charge order shows up in various forms depending on materials, and it overlaps other orders in the phase diagram. Because of this diversity and complexity it has been far from clear whether or not the charge order has a direct relevance to superconductivity and to the mysterious pseudogap phenomenon. However, the research development in the last few years has been successful in revealing universal aspects of the charge order and its fluctuations. It turns out that the charge order is compatible with superconductivity, and that its fluctuations have high onset temperature and high energy scale comparable to those of the pseudogap. The charge order correlations might be indispensable for creating the pseudogap and possibly generating high-$T_c$ superconductivity by collaborating with spin and pairing correlations.

## 1. Introduction

High-critical-temperature ($T_c$) superconductivity in the copper oxides (cuprates) emerges when the $CuO_2$ planes, which are insulating with a charge-transfer energy gap and with an antiferromagnetic (AF) order, are doped with charge carriers, holes or electrons. Static AF order disappears quickly as a function of doping, but the antiferromagnetism of the insulator survives in the superconductor in the form of dynamic spin fluctuations which are much stronger than in conventional metals.

More unique to the cuprates is the behavior observed below a temperature $T^*$ which is strongly dependent on the dopant concentrations ($p$) and is referred to as the "pseudogap" (PG) regime. It is characterized by a substantial suppression of the electronic density of states at low energies (the eponymous PG). The line $T^*(p)$ denotes the onset of a partial gap observed in spectroscopic data. The PG is the most mysterious regime in the generic $T$-$p$ phase diagram of hole-doped high-$T_c$ cuprates (shown in Fig. 1), and its understanding is considered to be essential for uncovering the high-$T_c$ mechanism. The energy scale of the PG is typically ~ 0.1 eV. The PG opens in the portion of the Fermi surface near the edges of the Brillouin zone (the antinode region of the $d$-wave superconducting order parameter), leaving behind the Fermi arc (centered at the $d$-wave nodes) where gapless and coherent quasiparticles reside. The opening of this anisotropic gap and the partial restoration of quasiparticle coherence from the totally incoherent 'strange metal' (SM) state above $T^*$ give rise to a dramatic change in various properties below $T^*$[1].

Another increasingly well documented feature of the PG regime is a tendency toward a variety of orders i.e. broken symmetries in addition to the gauge-symmetry broken superconductivity. Neutron scattering studies in the mid 1990s led to the experimental discovery of electronic "stripes" in $La_{1.6-x}Nd_{0.4}Sr_xCuO_4$[2], one of the La214 family including $La_{2-x}Ba_xCuO_4$ (LBCO) and $La_{2-x}Sr_xCuO_4$ (LSCO).

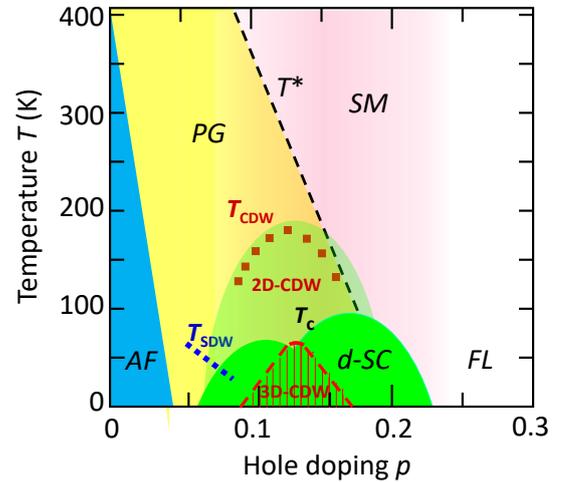

Fig. 1: Phase diagram of YBCO. AF, PG, d-SC, SM, and FL denote antiferromagnetic, pseudogap, $d$-wave superconducting, strange metal, and Fermi liquid phase, respectively. The onset of short-range CDW (SDW) order is indicated by squares. The 3D-CDW order in high magnetic fields is realized in the hatched region. In the dome-like area centered at $p \sim 0.12$, d-SC fluctuations are observed. Dynamic CDW correlations are present in the shadowed region.

A picture of the stripe order was based on the observation that doping an insulating antiferromagnet produces a tendency to phase separation which is frustrated by the long-range Coulomb interaction. To compromise, conducting stripe-like charge textures are formed in the stripe order[3]. It is characterized by an incommensurate array of charge stripes with incommensurate antiferromagnetic spin order in between, and is now found in the members of the LSCO family (La214 family) where a low-temperature-tetragonal (LTT) lattice deformation apparently acts as a pinning potential for the stripes. The cuprate stripes stay metallic and even superconduct at low temperatures.

For many years, the static stripe order had seemed to be confined to the La214 family. However, in early 2010s, charge ordering was discovered in underdoped YBa$_2$Cu$_3$O$_{7-y}$ (YBCO)[4-6] and Bi-[7,8] and Hg-based cuprates[9,10] (hereafter called non-La214 family). The charge order in a nanometer scale in Bi-based cuprate was already observed by the scanning-electron-tunneling spectroscopy (STS)[11-14]. The x-ray experiments found short-range incommensurate charge order that gradually onsets between 100 and 200K. A difference with the stripes in the La214 family is that there is no evidence of coincident static (or nearly static) spin order. Moreover, the variation of the charge order wavevector with doping in YBCO and Bi-based cuprates has the opposite sign of that in the La214 cuprates[15-17]. This difference may be connected with the presence of a spin gap in the non-La214 families which acts to suppress the spin order that is more prevalent in the La214 family[18].

An unusual relationship between the charge order and superconductivity has become apparent. In many classes of solids, such as transition-metal chalcogenides and organic compounds, a charge order (or a CDW phase) is observed in proximity to a superconducting (SC) phase[19]. Usually, the superconducting phase emerges by a suppression of CDW using chemical doping or applied pressure. Thus, CDW and SC are competing and mutually exclusive orders. The charge order in cuprates shows a similar competition with superconductivity. In a class of cuprates, the charge order weakens at the onset of superconductivity. However, a distinct feature of the cuprate charge order is its overlap with the SC order and SC fluctuation in the $T$-$p$ phase diagram. This indicates that charge order is not a simple competitor with SC, and it may coexist or be intertwined with SC.

The research development of the cuprate charge order before 2015 is summarized in, *e.g.* two reviews[1,20]. Basic questions on the cuprate charge orders that arose before 2015 were:

(i) Is the short-range CDW discovered in the non-La214 cuprates an order distinct from the charge stripe in La214 ?

(ii) Does the CDW form unidirectional like stripe order or bidirectional (checkerboard) order ? These two are difficult to distinguish in scattering experiments.

(iii) Is the CDW in the non-La214 static or dynamic ?

(iv) Does the real-space mechanism based on the strong electronic correlations as that for the stripe order work in the formation of the CDW ? Alternatively, is the momentum-space mechanism as that for the classical CDW systems more appropriate ?

(v) Do the charge correlations and/or the CDW order play a *secondary* role to the spin correlations in the cuprate superconductivity or is the CDW order a mere consequence of an hidden primary order that is responsible for the pseudogap?

In the present review the remarkable development in the research of charge order and its fluctuations in the cuprates after 2015 is overviewed[18,21] with focus on the interplay between charge order and superconductivity. The experimental discoveries over the past few years, thanks mainly to the rapid progress in the resonant x-ray scattering, have resolved most of the above questions and advanced our understanding of the charge order and its fluctuations in cuprates, while they forced us renewed considerations of high-$T_c$ superconductivity by taking into account spin, charge and pairing correlations on equal footing.

## 2. Diversity and Complexity of Cuprate Charge Orders
### 2.1 Stripe order in La-based 214 cuprates

The presence of charge order in the doped cuprates was theoretically predicted before the experimental discovery based on the models of strongly correlated electrons[22,23]. A theoretical picture of charge order was called a spin-charge stripe order composed of a periodic array of charge stripes and antiferromagnetically ordered spin stripes are between neighboring charge stripes. The periods of the spin and charge stripe array is incommensurate with the underlying lattice. The stripe order in real materials was initially found in a partially Nd-substituted LSCO, La$_{2-x-y}$Nd$_y$Sr$_x$CuO$_4$[2], and later La$_{2-x}$Ba$_x$CuO$_4$ (LBCO)[24]. In this La214 family the LTT lattice deformation apparently acts as a pinning potential for the charge stripes. Matching to the LTT structure, the orientation of the stripes changes by 90° as one moves from one layer to the next, forming three-dimensional (3D) spin-charge stripe order[2,25].

The spin stripe order onsets at lower temperature than the onset of the charge stripe order. It became clear that these stripes were different from the "classical" stripes in the other doped Mott-insulators[26,27]: the cuprate stripes stay metallic and even superconducting at low temperatures. While the stripe order suppresses bulk (3D) superconductivity, there is growing evidence that pairing, in a form distinct from the spatially uniform $d$-wave SC order, is taking place. This SC order can be viewed as a partially crystallized superconductor, formed from electron pairs (pair-density-wave, PDW)[28-31]. The cuprate-stripe is thus a unique and most extreme case of charge, spin, and electron-pair intertwined orders.

*Two-dimensional superconductivity (2D-SC)*

The evidence that the charge stripes are internally superconducting has been provided by various experiments[18,29]. Given that the stripe orientation varies 90° on moving from one layer to the next, and if SC order between one stripe and the next is antiphase within a layer, the Josephson coupling between layers is completely frustrated[28,30]. The frustration of the interlayer Josephson coupling was experimentally confirmed by the disappearance of the Josephson plasma edge in the $c$-axis infrared reflectivity spectrum [32]. This interlayer decoupling isolates each superconducting CuO$_2$ plane and thus a two-dimensional (2D) superconducting state emerges in the stripe-ordered state.

For LBCO with $x$ = 1/8 the onset of 2D pairing correlation within the CuO$_2$ plane coincides with the onset of the spin-stripe order ($T_{SO}$ ~ 40 K)[33]. At $T_{SO}$ a steep drop of the in-plane resistivity ($\rho_{ab}$) is observed whereas the c-axis resistivity ($\rho_c$) continues to increase, but the in-plane resistivity does not reach zero (see Fig. 2(a)). With further cooling below ~ 16 K, $\rho_{ab}$ becomes vanishingly small but $\rho_c$ is still finite, evidencing that the pairs establish a long-range coherence in a plane - 2D superconductivity order - through a Berezinskii-Kosterlitz-Thouless (BKT) transition at $T_{BKT}$ ~ 16 K. The pronounced nonlinear current-voltage characteristics observed below 16 K yields further support the 2D-SC[34,35]. Note that this stripe-ordered state provides an unprecedented playground of the BKT transition and 2D-SC realized in bulk crystals at very high temperature and a very wide temperature range, as compared to those in metal thin films and ion-gated superconducting monolayers[36].

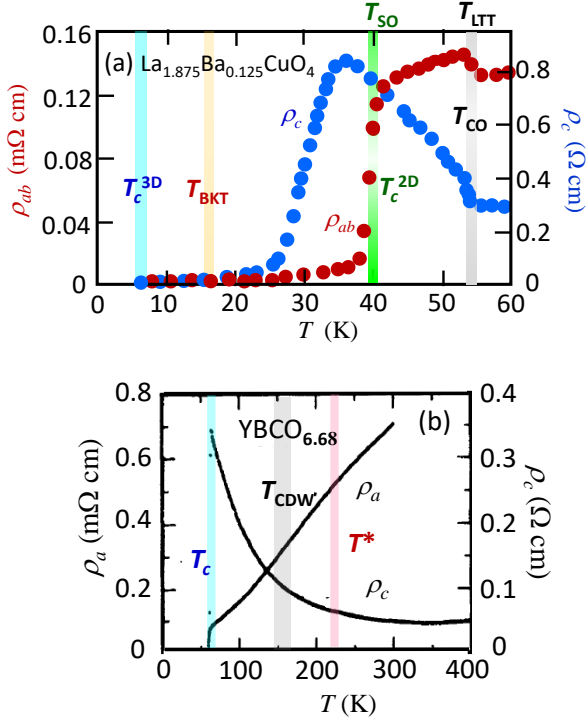

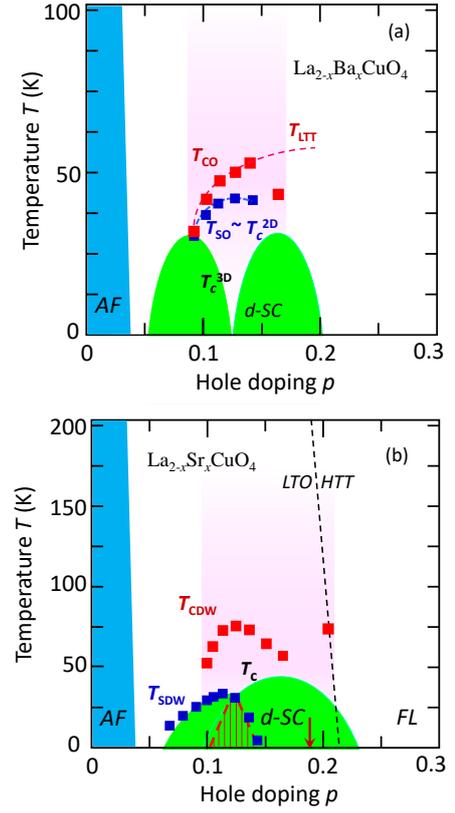

Fig. 2: In-plane ($\rho_{ab}$ or $\rho_a$) and c-axis resistivity ($\rho_c$) plotted as a function of temperature for (a) $La_{2-x}Ba_xCuO_4$, adapted from ref. 35 and (b) $YBa_2Cu_3O_{6.68}$ with $p \sim 1/8$. For LBCO, the onset temperature of charge-stripe order, $T_{CO}$, coincides with the LTT structural transition temperature ($T_{LTT}$), and at the onset of spin-stripe order, $T_{SO}$, superconducting correlations within each $CuO_2$ plane starts to develop ($T_c^{2D}$). The BKT transition and the bulk 3D superconducting transition temperature ($T_c^{3D}$) are also indicated. For YBCO, the onset of the short-range CDW ($T_{CDW}$) in addition to two characteristic temperatures, pseudogap temperature $T^*$ and superconducting transition temperaature $T_c$, are indicated by the vertical lines.

Fig. 3: (a) Phase diagram of $La_{2-x}Ba_xCuO_4$. The onset of charge (spin)-stripe order is indicated by squares. The dashed line is the doping dependent LTT structural transition. In the shadowed region a precursory or fluctuating stripe order is observed. (b) Phase diagram of $La_{2-x}Sr_xCuO_4$. The squares indicate the onset of incommensurate short-range charge (CDW) order and that of spin (SDW) order. The data point of $T_{CDW}$ at $x = 0.21$ is from ref. 47. The dashed line shows the high-temperature-tetragonal (HTT)-low-temperature-orthorhombic (LTO) structural transition, and the arrow at $x = 0.18$ indicates the Sr content at which the Fermi surface topology changes due to the Fermi level crossing the van Hove singularity. The static stripe-like order presumably takes place in the hatched area. The fluctuating charge order is observed in the shadowed region.

The coexisting spin stripe order is a prerequisite for realizing the supposed PDW state with antiphase SC order between neighboring charge stripes within a layer. In an array of antiphase charge stripes the pair wavefunction can get zero where the amplitude of spin stripe is maximum which matches the empirical rule that static SC order avoids spatial overlap with static AF order[37]. This might be the reason why the onset of 2D-SC coincides with $T_{SO}$.

### Three-dimensional superconductivity (3D-SC)

To achieve bulk (3D) superconductivity, it is necessary to lock together the SC phases of the $CuO_2$ planes via interlayer Josephson coupling. When the 2D superconducting correlations within the plane sufficiently develop, that happens below the onset of the spin-stripe order ($T_{SO}$) in LBCO, even a small interlayer Josephson coupling is enough to realize 3D superconductivity. The stripe order is strongest at $x = 0.125$ for LBCO (and Nd-substituted LSCO), and hence the frustration of the interlayer Josephson coupling is strongest. As a consequence, 3D bulk superconductivity occurs only at temperature $T_c \sim 5$ K, as signalled by the zero resistivity and substantial diamagnetism[33,35]. When $x$ in LBCO is changed to slightly smaller or larger value than 1/8, the stripe order gets weaker, and the onset of 3D superconductivity ($T_c^{3D}$) rapidly increases (see the phase diagram of LBCO in Fig. 3(a)) – $T_c^{3D}$ rises to 14 K for $x = 0.115$ and 16 K for $x = 0.135$. Consequently, the $T_c^{3D}$ vs $x$ curve has a deep dip at $x = 1/8$, forming two $T_c$-domes.

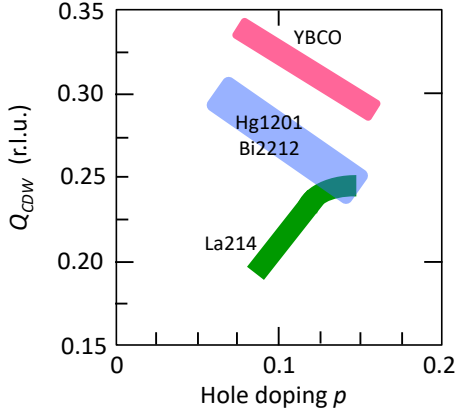

Fig. 4: A rough sketch of the doping dependence of the magnitude of the charge order (CDW) wavevector for four representative cuprates.

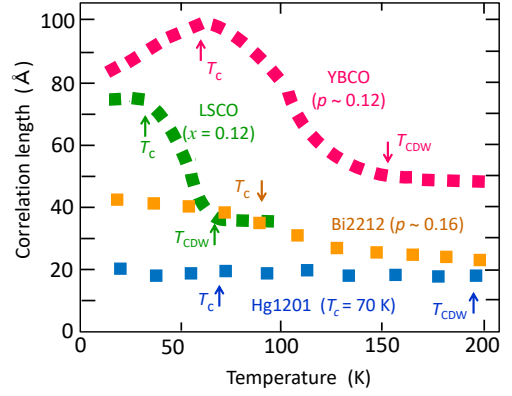

Fig. 5: Temperature evolution of the in-plane CDW correlation length for the four cuprates with doping level at which the strength of the CDW order is maximal, except for Bi2212 with near optimal doping. The onset of 2D-CDW ($T_{CDW}$) and SC ($T_c$) are indicated by arrows. $T_{CDW}$ is not well-defined in Bi2212.

A compressive uniaxial strain, induced either by external uniaxial pressure[38] or by lattice mismatch between LBCO films and substrate materials[39], dramatically changes the competition between stripe order and 3D SC. Under these strains, $T_c^{3D}$ is found to increase to a temperature comparable to the onset of spin-stripe order or the onset of 2D SC correlation ($T_c^{2D} \approx T_{SO} \sim 40$ K). In particular, $T_c^{3D}$ of the strained LBCO thin films rises over an entire $x$ range up to the values higher than $T_c$ in the unstrained bulk LBCO, and as a result, the dip at $x = 1/8$ disappears. The LTT structure is no more observed in the strained LBCO films. The absence of the LTT structure could weaken the lattice-pinning of the stripe order, thereby restoring the interlayer phase coherence.

**2.2 CDW orders in non-La214 cuprates**

After 2011 charge ordering was discovered in various non-La214 families in the underdoped regime by resonant-x-ray-scattering (RXS). Non-La214 families include YBa$_2$Cu$_3$O$_{7-y}$ (YBCO), Bi-based cuprates, Bi$_2$Sr$_2$CaCu$_2$O$_{8+\delta}$ (Bi2212) and Bi$_2$Sr$_2$CuO$_{6+\delta}$ (Bi2201), a Hg-based cuprate, HgBa$_2$CuO$_{4+\delta}$ (Hg1201), and also electron-doped cuprates, (Nd,Pr)$_{2-x}$Ce$_x$CuO$_{4-y}$[40]. From the x-ray diffraction and scattering experiments, it was found that these are all short-range incommensurate charge orders with ordering vectors ($\boldsymbol{Q}_{CDW}$) always along the Cu-O bond directions as in the case of the stripe order in La214. Different from the stripe order, the charge order (hereafter dubbed CDW) is confined within a CuO$_2$ plane, i.e., two-dimensional (2D). The 2D charge order (2D-CDW) gradually onsets at temperature ($T_{CDW}$) between 100 and 200 K in YBCO. As regards to doping dependence, the onset temperature, though sometimes difficult to unambiguously define, appears to form another dome as a function of the doped hole density $p$ with a maximum at $p \sim 1/8$ except for Hg1201. The dome starts from $p \sim 0.08$ and terminates at the upper doping end near $p = 0.18$. As $T_{CDW}$ is always lower than the PG temperature $T^*$, the CDW dome exists within the pseudogap (PG) region. Experimental probes with longer time scales, such as nuclear-magnetic-resonance (NMR), have confirmed that the short-range charge order is almost static[41], presumably due to *pinning* of correlated charge fluctuations by lattice disorder necessarily existing outside the CuO$_2$ planes as dopants in most of the cuprates.

A difference from the stripes in the La214 family is that there is no evidence of coincident static (or quasistatic) incommensurate spin order (or spin-density-wave, SDW), instead a gap opens in the spin excitation spectrum. Moreover, the variation of the charge order wavevector ($\boldsymbol{Q}_{CDW}$) with doping in non-La214 cuprates has the sign opposite to that in the striped La214 (the wavevector ($\boldsymbol{Q}_{CO}$) of the charge stripes) as schematically depicted in Fig. 4. In the latter, $Q_{CO}$ increases with doping as expected in a real space picture, i.e. a distance between charge stripes. To the contrary, $Q_{CDW}$ in the former decreases which would be expected from a momentum space picture involving vectors spanning the Fermi surface, specifically, a distance between the tips of the Fermi arc[20]. This is analogous to the Fermi surface nesting vectors in classical CDW systems. Because of this, charge order in non-La214 is frequently referred to CDW.

When the SDW is dynamic with a spin gap, $Q_{CDW}$ is expected to unlock from the spin correlations, in contrast to the stripe order in the La214 cuprates where the charge-order wavevector $Q_{CO}$ is locked to that of the spin order $Q_{SO}$ to satisfy the relation $Q_{CO} = 2Q_{SO}$[2], similar to the case of 'classical' stripe order in, e.g., hole doped nickelate La$_{2-x}$Sr$_x$NiO$_4$[26].

***Two-dimensional CDW order (2D-CDW)***

The onset of 2D-CDW ($T_{CDW}$) is signaled by the appearance of a peak at $\boldsymbol{Q}_{CDW}$ in RXS. The peak is always broad, indicating that the 2D-CDW is a short-range order. The in-plane correlation length ($\xi$) estimated from the peak width is about a few CDW periods, ~ 15 – 25 Å in Bi- and Hg-based cuprates and also electron-doped cuprates, whereas $\xi$ of YBCO is considerably large, ~ 40 Å[42]. The larger $\xi$ in YBCO was ascribed to minimal chemical disorder in this cuprate, since the dopant oxygen atoms in the Cu-O chains are ordered on the length scale of a few hundreds Å.

For the 2D-CDW state of YBCO, $\xi$ increases strongly upon cooling from $T_{CDW}$[42] (see Fig. 5), seemingly toward a truly long-range order at low temperature, if not a superconducting order sets on at $T_c$. By contrast, for Hg1201 $\xi$ remains small even at low temperatures[9,10], which may be ascribed to stronger lattice disorder[21]. However, the small and temperature independent $\xi$ in Hg1201 is puzzling considering that $\xi$ in more disordered Bi2212 (as described in 3.2) is equally small but shows a weak increase upon cooling, and that the effect of disorder on the SC order or the SC $T_c$ is minimal in Hg1201 where the chemical disorder due to excess oxygen atoms is only in the HgO blocks, located fairly distant from the CuO$_2$ planes. This may indicate that the same disorder differently affects CDW and SC order – the CDW is more sensitive to disorder than the SC order[43] or that the short $\xi$ is inherent to Hg1201. The disorder in high-$T_c$ cuprates is known either to locally suppress SC or to have no substantial influence on SC depending on the type and the location of disorder[44]. For CDW, disorder tends to disrupt a long-range CDW correlations on one hand, and to locally nucleates or stabilizes CDW order by pinning fluctuating CDW on the other. To see real effect of disorder on CDW, it is necessary to experimentally cover more cuprate families including the stoichiometric YBa$_2$Cu$_4$O$_8$.

As regards the doping dependence, the 2D CDW order is observed for most cuprate families in the range $0.08 < p < 0.18$ with its maximum onset $T_{CDW}$ at $p \sim 1/8$. Hg1201 is an only exception in which a maximum appears to be at lower $p \sim 0.095$[10], but the determination of the correct doping level is necessary for Hg1201.

While the CDW order terminates at $p \sim 0.18$ in the bilayer Bi2212 as observed by STS[45], a 're-entrant' CDW without weakening in strength was observed by RXS measurement on a nominally heavily overdoped single-layer Bi2201[46]. Something special may happen in Bi2201. One possibility is that at this doping level the Fermi level passes through a van Hove singularity and the electronic density of states is enhanced to strengthen CDW correlations. This seems a case also with LSCO ($x = 0.21$)[47], the phase diagram of which is shown in Fig. 3(b). Another is a broad distribution of the local hole density in the sample of Bi2201. The doping levels of Bi2201 are quite uncertain. The widely used empirical relation between $T_c$ and hole density $p$ is not applicable to Bi2201.

## LSCO

A representative member of the La214 family, La$_{2-x}$Sr$_x$CuO$_4$ (LSCO) is unique or rather exceptional in the perspective of stripe order. The low-energy spin and charge fluctuation spectra very much resemble those for other members showing static stripe order. However, an accompanying static (or quasistatic) spin stripe order, incommensurate SDW, is observed only below $x \sim 0.14$[48,49], and there is no LTT structure at any $x$. Consequently, the quasistatic stripe order in LSCO never attains three dimensionality. Above this concentration, a spin gap, instead of spin stripe order, develops[50-52]. A rapid increase of $\xi$ below the onset of 2D-CDW (or charge stripe) order, is also observed for LSCO [46] (Fig. 5), suggestive of a charge ordering similar to that in YBCO. In this respect, the moderately and highly hole-doped LSCO appears to share features with the non-La214 family.

## Three-dimensional CDW order (3D-CDW)

The three-dimensional CDW (3D-CDW) order was discovered by x-ray experiments on underdoped YBCO either in high magnetic field (> 15 Tesla) applied perpendicular to the CuO$_2$ planes (along the crystal $c$-axis)[53-57] or under uniaxial strain (~ 1 % compression) applied parallel to the planes[58,59]. This order is a new type of CDW order confirmed by appearance of sharp scattering peaks centered at the same in-plane $q = Q_{CDW}$ as that of the 2D-CDW but at integer values of the momentum transfer in the $c$-axis direction. This evidences a 3D nature distinct from the 2D-CDW observed at zero magnetic field and zero strain. Like the stripe order in LBCO with $x = 1/8$, this CDW is a truly static long-range order with correlation lengths exceeding 300 Å in the plane and about 50 Å between layers, and certainly identical to that previously observed by the nuclear-magnetic-resonance (NMR) experiment in high magnetic fields[4,41]. Both x-ray scattering and NMR indicate that the magnetic-field-induced 3D-CDW is uniaxial in the plane, reminiscent of the stripe order in La214 cuprates.

YBCO exhibits quantum oscillations (QOs) in magnetic fields of typically larger than 20 Tesla where the long-range 3D-CDW is stabilized. The Fermi surface inferred from QOs is a small electron pockets in contrast to the hole Fermi surface (or Fermi arc) at zero magnetic field. It is thought that this Fermi-surface-reconstruction (FSR) is caused by the 3D CDW, but several questions about 3D-CDW remain open:

1) The electron Fermi surface inferred from QOs better matches the models with bidirectional CDW than unidirectional one[60]. This is also the case with Hg1201 which exhibits QOs at low temperatures[61].

2) No experimental result has so far been reported on the development of 3D-CDW for Hg1201. So, a link between QOs and 3D-CDW is not clear at present. Then, another question is whether the 3D-CDW order is specific to YBCO or exists universally in the cuprates.

3) The 3D-CDW induced by uniaxial strain in YBCO is also unidirectional in the plane which appears only under the strain along the $a$-axis and has the CDW wavevector parallel to the $b$-axis (in the CuO chain direction in YBCO)[59]. It is not understood yet why the $b$-axis strain does not induce a 3D-CDW with an $a$-axis wavevector. Perhaps, the presence of the CuO chains might be a cause of this anisotropy, or the

*b*-axis strain in the experiment is not strong enough to induce an *a*-axis CDW.

4) At present, FSR is not reported in the strain-induced 3D-CDW state either by quantum oscillations or by angle-resolved-photoemission-spectroscopy (ARPES). Again, the conclusion that the 3D-CDW is an only cause of FSR or QOs is not justified at present.

## 2.3 Relationship between the stripe order in La214 and CDW orders in non-La214

The so far discovered CDW orders appear to be distinct from the stripe order. One of the differences from the stripe order is that there is no coincident static (or nearly static) spin order. Another difference is in the variation of the CDW wavevector with doping as displayed in Fig. 4. Whereas in the case of stripe order, this wavevector, $Q_{CO}$, increases with doping as expected in a real space picture, in the CDW order, $Q_{CDW}$ decreases as would be expected from a momentum space picture. As described before, these differences are likely connected with differences in the spin behavior.

The recent experimental results apparently challenge this view:

(1) As will be discussed in the next section, the CDW correlations at temperatures higher than $T_{CDW}$ (or $T_{CO}$) appear quite similar in both families[62-66].

(2) The observed pattern of charge density modulations by scanning-tunneling-spectroscopy (STS) for Bi2212 is unidirectional like the stripe[67-69], and also the 3D CDW of YBCO in high magnetic fields[4,41,53-56] or uniaxial stress[58,59] (and probably 2D CDW) is uniaxial.

(3) The charge order wavevector $Q_{CO}$ of the charge stripe order in LBCO, which is locked to that ($Q_{SO}$) of the spin stripe with $Q_{CO} = 2Q_{SO}$ at low temperatures, becomes unlocked upon heating above $T_{CO}$ = 55 K[63,70]. $Q_{CO}$ of the stripe order weakly increases with raising temperature from 0.24 r.l.u. (reciprocal lattice unit) below $T_{CO}$ to 0.27 r.l.u. above $T_{CO}$[64], while the magnitude of $Q_{SO}$ strongly decreases from $2Q_{SO}$ = 0.24 at low $T$ to ~ 0.18 r.l.u. at high $T$.

(4) In the case of the CDW in non-La214, the magnitude of the CDW wavevector $Q_{CDW}$ is typically ~ 0.3 r.l.u. This value turns out to decrease upon heating towards 0.25 r.l.u. for Bi2212[71]. The high-$T$ value of $Q_{CDW}$ is close to that of typical $Q_{CO}$ for the stripe order, suggesting a universal CDW period over all cuprates. This may be compatible with the STS observation of the locally commensurate CDW with $Q_{CDW}$ exactly 1/4[72].

All together, a plausible scenario would be that many CDW states with different $Q_{CDW}$'s (both amplitude and direction as will be described in the next section) can exist with only small energy differences, and they show up as dynamic CDW correlations almost ubiquitously observed at high temperatures. At low temperatures, $Q_{CDW}$ in La214 would be determined by coupling between CDW and SDW that organizes well-correlated stripe order. On the other hand,

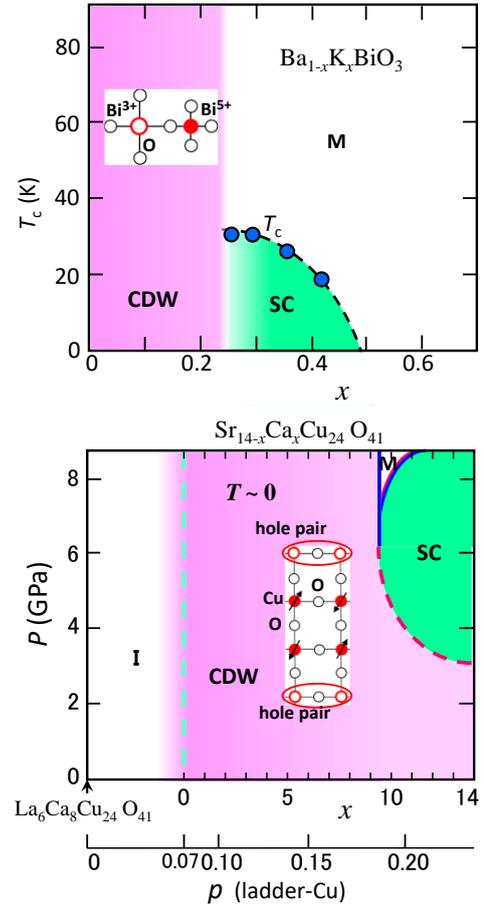

Fig. 6 (a): Phase diagram of $Ba_{1-x}K_xBiO_3$. The CDW order is quite stable and persists to very high temperature even at the boundary between CDW and superconducting phase. The inset shows a local correlation of charge disproportionation of neighboring Bi ions and associated oxygen breathing-mode distortions. (b) Pressure-doping phase diagram of the hole-doped two-leg ladder cuprate. The undoped ($p$ = 0) two-leg ladders is realized in $La_6Ca_8Cu_{24}O_{41}$ which is a charge-transfer insulator. The charge order appears at doping level $p > 0.05$ (the doping level $p$ in the figure is estimated from the low-energy optical spectral weight[78], but there is a debate on the value of $p$[80]). At low pressure CDW is robust over the entire doping range and the compound remains insulating. When the the two-leg ladders are highly doped with holes, superconductivity (SC) emerges by applying high pressure ($P$ > 3 GPa). At higher pressure the compound is metallic without SC. The inset is a sketch of a supposed charge order with hole pairs on the rung periodically align along the leg direction.

in non-La214 this mechanism does not work due to the presence of a spin gap and the absence of the LTT lattice structure. In this case, other material-specific details may be relevant for determining $Q_{CDW}$, for example, Fermi surface topology or the position of a van Hove singularity relative to the Fermi level.

There has been a debate on the mechanism of the CDW formation in non-La214 cuprates. The stripe order in La214 nicely fits the real-space picture, so that its formation is no doubt driven by the strong coupling mechanisms originating from a compromise between local spin superexchange interaction and kinetic energy. On the other hand, the doping dependence of $Q_{CDW}$ in non-La214 seemed to favor a weak coupling Fermi surface nesting picture. However, the Fermi-surface nesting picture is challenged by the ARPES study of underdoped Bi2212 which indicates no gap opening at the tips of the Fermi arc in the 2D-CDW state[73]. As will be discussed in the next, the experiments in the last three years obviously point toward the strong coupling picture.

**2.4 Interplay between CDW and superconducting orders**

It would be worth referring the two CDW systems for consideration of the mechanisms of the CDW formation and of the relationship between CDW and superconductivity in cuprates; the perovskite bismuth oxides $Ba_{1-x}K_xBiO_3$[74] and the two-leg ladder cuprate $Sr_{14-x}Ca_xCu_{24}O_{41}$ (14-24-41)[75]. Both share some similarity to the high-$T_c$ cuprates; the phase diagrams for two systems are shown in Figs. 6(a) and (b).

The parent bismuthate $BaBiO_3$ is a three-dimensional (3D) analog to the 2D cuprates in that it is a half-filled that single-band system composed of strongly hybridized Bi6$s$ and O2$p$ orbitals in the weak correlation limit. An extremely robust 3D-CDW, that is stabilized by the frozen breathing phonon mode, persists to 1000 K or even higher temperatures is formed in this oxide[76]. The SC order with maximum $T_c$ of ~ 30 K appears when chemical doping proceeds to fairly high levels. The driving force of the CDW formation is a quite local correlation as in the case of the cuprate CDW. In this case the local correlation is a tendency for charge disproportionation of neighboring Bi atoms into $Bi^{5+}$-$Bi^{3+}$ pair (or Bi-O bond disproportionation[77]), a kind of local attractive interaction between electrons, that arises from the strong relativistic effect ("mass-correction" term in the special relativity) of the heavy Bi atom.

The 14-24-41 cuprate is a 1D analog of the 2D cuprates, and can be doped with holes[78]. Charge order, presumably an ordering of doped hole pairs, occurs on the doped two-leg ladders over a wide doping range[79,80]. A hole-pair formation and its ordering are thought to be driven by the local superexchange interaction ($J$) between neighboring Cu atoms[81]. The pair formation via $J$ appears to have a resemblance to the stripe order in the La-family of cuprates[18]. In this ladder cuprate, SC emerges by applying high pressure for highly hole doped compounds with $T_c$ ~ 12 K[75,82]. The charge order of hole pairs is a sort of PDW, but the superconducting state realized under pressure is not necessarily a PDW superfluid. The superconductivity in 14-24-41 appears to result from a melt of the charge order and a simultaneous increase in the dimensionality from 1D to 2D[82].

In both oxides, as in the classical CDW systems, the CDW phase lies in proximity to the SC phase. CDW and SC are competing and mutually exclusive orders. It seems that the relation between CDW and SC is similar to that between AF and SC in the two-dimensional cuprates.
The charge order in cuprates shows a similar competition with superconductivity; the superconducting $T_c$ is reduced around the doping at which the 2D-CDW is strongest, and conversely the quasielastic CDW signal of RXS weakens at the onset of superconductivity. However, a distinct feature of the cuprate charge order is its overlap with the SC phase in the $T$-$p$ phase diagram. In particular, as shown in Fig. 1, the $T$-$p$ region of the 2D-CDW is nearly coincident with that of the superconducting fluctuations characterized by the sizable Nernst effect and diamagnetic response[1]. This suggests that charge order in cuprates is not a simple competitor with SC.

There are now growing evidences that *the CDW is compatible with superconducting pairing correlation* in the high-$T_c$ superconducting cuprates. For the extensively studied cuprate families, YBCO, Bi2212[65], and LSCO with $x > 0.14$, the intensity of the 2D-CDW signal radically drops below $T_c$, but the 2D-CDW is not completely suppressed even at temperatures well below $T_c$, suggestive of coexisting CDW and SC. For others, Bi2201, LSCO with $x < 0.14$, and Hg1201, the 2D-CDW intensity does not show a significant drop below $T_c$. In these families the CDW and SC orders apparently coexist without any appreciable the influence on each other. A common feature of the former two is their low superfluid density ($\rho_s$) and low $T_c$[83] due to strong disorder in Bi2201 and to stronger tendency to spin stripe order in LSCO. The electron-doped cuprates turn out to belong to this class[40]. By contrast, the SC order in Hg1201 is strongest among the known cuprates, suggesting that too strong or too weak SC order has weaker influence on the CDW order.

Unlike the 2D-CDW, the magnetic-field-induced 3D-CDW (and also the uniaxial strain-induced 3D-CDW) in YBCO appear to be mutually exclusive, a typical manifestation of thermodynamic competition between two 3D long-range ordered phases[56,84]. However, the interplay between 3D-CDW and SC is far from simple. Uniaxial strain of ~ 1% compression along the a-axis induces 3D-CDW at 75 K for an underdoped YBCO ($p$ ~ 0.12), but SC is not completely suppressed with $T_c$ = 65 K at zero strain being reduced to ~ 48 K[59]. Upon the onset of SC the signal of the 3D-CDW radically diminishes. Even more puzzling is the magnetic-field-induced 3D-CDW. It Whether the SC order or SC correlation survives at low temperatures well below the onset of the 3D-CDW is a matter of debate. This issue is linked to the origin of the QOs and hence the FSR.

Earlier resistivity measurement on underdoped YBCO in magnetic fields suggested that the upper critical field $H_{c2}$ at $T = 0$ might be 20 - 25 Tesla[85], considerably lower than that deduced, for example, from the very short SC coherence length. Such a low field coincides with the field above which QOs are observed, and consequently FSR and QOs were thought to be the normal-state properties associated with the long-range 3D-CDW order. However, the careful resistivity measurement recently performed[86] has demonstrated that the finite resistivity at fields higher than the nominal $H_{c2}$ is due to too large electric current used for the resistivity measurement in the previous study, and that the actual high-field and low-temperature state is a very fragile non-dissipative state. Most likely it is this state that exhibits QOs and hence FRS, and coexists with 3D-CDW. The true nature of this exotic state, named a quantum vortex matter, is not fully elucidated, but a PDW state is proposed as a possible candidate[86,87].

A robust SC correlation remaining in the magnetic-field-induced 3D-CDW phase is also speculated by the high magnetic field study of specific heat and Knight shift[56]. The results suggest that the two mutually exclusive orders, 3D-CDW and 3D-SC, establish some form of cooperation in order to coexist at low temperature, and a PDW state could be a candidate of such a collaboration[31].

Experiments that directly address the question are difficult at present, but there are some hints in the experiments performed in magnetic fields. One is the STS observation of PDW within vortex halos for Bi2212[88], although it is not clear if 3D-CDW is stabilized in the vortex halo region. Another hint is from a result of the infrared optical study of the c-axis Josephson plasma under magnetic fields for an underdoped YBCO with $p$ near 1/8[89]. The applied magnetic fields ($B \leq 7$ T) are considerably lower than the critical field ($\sim 15$ T) to induce 3D-CDW, the Josephson plasma frequency, a measure of the interlayer Josephson coupling strength, is found to show a rapid red shift with magnetic field – the inter-bilayer Josephson coupling strength at $B = 7$ T is less than half of that at $B = 0$, much bigger than expected from the dephasing effect due to vortex meandering. This finding suggests a possibility that the magnetic-field-induced 3D-CDW state might coexist with an inter-bilayer-phase-decoupled SC state, possible PDW, as in the case of LBCO. However, a question remains if a supposed PDW state without coexisting SDW order can frustrate interlayer phase coherence

## 3. CDW Fluctuations

For layered transition-metal dichalcogenides such as 2H-TaSe$_2$, one of the classical CDW systems, it was argued that CDW with short correlation length should be subject to strong CDW fluctuations[90]. Considering the very short correlation lengths, it is natural to expect that the 2D-CDW in the cuprates with strongly correlated electrons shows robust and unique fluctuations in a form of dynamic CDW correlations and their associated collective charge excitations. Recent advanced theoretical calculations based on the Hubbard model have found fluctuating stripe order at a significantly high temperature[91]. It was also theoretically predicted that stripe fluctuations would cause a softening of relevant phonon modes similar to the Kohn anomaly in classical CDW systems[92].

### 3.1 CDW fluctuations - Dynamic CDW correlations

The recent extensive study of the resonant-inelastic-x-ray-scattering (RIXS) of NdBa$_2$Cu$_3$O$_{7-y}$ (and YBCO) have attributed the observed broad peak, being centered near $Q_{CDW}$, to charge density fluctuations signal arising from dynamic CDW correlations[62]. The dynamic CDW correlations are found to exist over a wide region of the $T$–$p$ phase diagram (Fig. 1). They persist up to temperatures well above $T_{CDW}$ and even above the PG temperature $T^*$ are thought to be precursor to the low-temperature quasistatic 2D-CDW. The precursor CDW (PCDW) is also two-dimensional, and the width of this peak is very broad and nearly temperature ($T$)-independent. The in-plane CDW correlation length $\xi$, estimated from the half-width at half-maximum, is extremely short, only one CDW period or less and $T$-independent. Along the doping-axis, they are observed from the lowest end ($p \sim 0.085$) of the quasi-static CDW to even the overdoped region ($p \sim 0.2$). The quasistatic CDW signal emerges as a narrow peak below $T_{CDW}$ while the broad peak is unchanged. The correlation length of the quasistatic 2D-CDW rapidly grows with cooling below $T_{CDW}$ but the growth gets slower toward the onset of SC. A typical temperature dependence of $\xi$ in YBCO with $p \sim 1/8$ is shown in Fig. 5. Notably, the integrated scattering intensity (spectral weight) of the broad peak is by about an order of magnitude larger than that of the narrow peak, indicating that only a small fraction of the total CDW correlations condenses into the static (quasistatic) low-temperature 2D-CDW order due possibly to defect pinning. Below the SC onset, no discernible change is observed for the broad peak whereas the intensity of the narrow peak decreases due to the competition with the SC order.

An important question on the RIXS evidence of dynamic CDW correlations (or PCDW) is whether they are ubiquitous in all the cuprate families. The answer to this question may be yes[93]. At present, similar but not exactly the same high-$T$ dynamic CDW correlations have been observed for Hg1201[66] and electron-doped cuprates[94]. The intensity of the quasielastic peak in Bi2212 appears to trace similar $T$-evolution to that in YBCO[65], but a high-$T$ region of dynamic CDW correlations where $\xi$ is temperature independent is not clearly identified. Hg1201 is again peculiar in this respect. $\xi$ of Hg1201 is very short ($\sim 20$ Å) and temperature independent even below $T_{CDW}$ ($\sim 200$ K), at which the quasistatic CDW order starts to develop, down to temperatures well below $T_c$[10], as if CDW in Hg1201 were fluctuating all the way down to the lowest temperature without onset of quasistatic CDW order.

The presence of fluctuating charge stripe has been confirmed for the La214 cuprate family (see Figs. 3(a) and (b)). In LBCO with $x = 1/8$, static stripe order is absent above $T_{CO} = 55$ K ($= T_{LTT}$), instead the charge-stripe fluctuations with dynamic stripe correlations have now been observed by RIXS at temperatures above 55 K[63,64]. They show up in RIXS as a broad peak centered near $Q_{CO}$ via the x-ray resonant process, and this diffuse high-$T$ scattering is found to comprise 2D $q$- and energy-integrated spectral weight by about 7 times larger than the sharp low-$T$ CDW peak, similar to the PCDW in the YBCO family. A difference from YBCO is that the static stripe order below $T_{CO}$ (or $T_{SO}$) acquires three dimensionality, while the CDW in YBCO is essentially 2D below $T_{CDW}$.

As displayed in Fig. 5, the $T$-evolution of $\xi$ in LSCO with $x \sim 1/8$ is similar to that in YBCO, and $\xi$ becomes $T$ independent above 70 K[47], typical of the PCDW region. As depicted in Fig. 3(b), fluctuating or incipient charge order exists over a broad range of doping $p$ ($= x$), $0.11 < x \leq 0.21$, not restricted in the underdoped region but extended to overdoped region across a critical doping $x_c \approx 0.18$ at which Fermi surface topology changes[47,95,96]. Apparently, this is against the Fermi-surface nesting scenario, and also against the presence of a putative quantum critical point associated with charge order.

Given that PCDW is ubiquitous across different families of cuprates[93], PCDW or dynamic charge fluctuations might serve as the 'seed' of the 2D-CDW that couples strongly to different types of correlations at low temperatures. In this context, cuprate CDWs may arise from the same underlying instability despite their apparently different low-temperature ordering wavevectors.

*Dynamic quasi-circular charge correlations*

A very recent energy-integrated resonant x-ray scattering (EI-RXS) study of Bi2212 has revealed a quasi-circular in-plane scattering pattern[97]. A scattering peak is observed in any in-plane direction at nearly the same $q$ close to $Q_{CDW}$ in the quasi-static 2D-CDW state in which the quasi-elastic peak is observed only in the Cu-O bond directions. The observed isotropic peak appears via inelastic channels in the mid-infrared energy range between 0.1 and 0.9 eV, signalling the presence of dynamic quasi-circular charge correlations associated with CDW. It is tempting to relate this result to the PCDW observed by RIXS at temperatures above $T_{CDW}$ and to conclude that the PCDW is unlocked in its direction or equivalently, uniaxial short-range CDW domains are fluctuating into any direction in the PCDW regime. However, this is an open question to be resolved by future angular dependent RIXS measurements on various cuprate families. At present, similar dynamic circular CDW correlations have only been observed for the electron-doped cuprates, T'-$Nd_2CuO_{4-y}$, in the heavily uunderdoped region ($n \sim 0.07$)[98].

The scattering peak arising from PCDW is so broad that PCDW equally affect almost all states on the Fermi surface, resulting in an essentially isotropic scattering that is a distinguished feature of the strange metal (SM)[1]. The quasi-circular correlations further provide a large circular manifold in momentum space which, together with the broad peak width, might be the reason for the much larger spectral weight of PCDW than that of the quasistatic 2D-CDW order and the long-range 3D-CDW at lower temperatures.

**3.2 Collective charge excitations and CDW gap**

CDW is a translational symmetry breaking order, therefore, in principle, there exist (i) an order parameter associated with CDW order, a CDW gap in this case, (ii) a characteristic collective excitations, and (iii) topological defects[99].

*Topological defects*

Topological defects associated with the low-temperature CDW order were directly observed for Bi2212 by STS[100]. They exist with a fairly high density that make the CDW correlation length $\xi$ very short (~ 20 Å). The positions of the topological defects are just mapped on the location of excess oxygen atoms in the $Bi_2O_2$ blocks[101], and also on the Sr sites in the $Sr_2O_2$ blocks in which Sr atoms are replaced by Bi stoms due to inter-site mixing[102] - $Bi^{3+}$ ions in the Sr sites tend to attract the excess $O^{2-}$ ions in the BiO layers. The Bi-Sr intersite mixing is thought to be a major source of disorder in real Bi2212 (and also Bi2201) crystals.

*Collective excitations*

As the collective excitations are fluctuations of charge density, in addition to a gapless Nambu-Goldstone mode (phase mode) emanating from $Q_{CDW}$, a gapped collective mode (amplitude mode) appears. In the case of stripe order, the phase mode corresponds to transverse meandering motion of charge stripes[103]. In real materials the phase mode is also gapped due to the pinning of CDW by defects and impurities in the underlying lattice. RIXS and EELS (low-energy-electron-loss-spectroscopy) are suitable for the observation of these modes as momentum ($q$)- and energy ($\omega$)-resolved probes.

The collective excitations in cuprates are also complex in their own and diverse among different cuprate families. Other than CDW excitations, there should be low-energy collective modes, paramagnons, acoustic plasmons, and phonons associated with spin, charge, and lattice degrees of freedom which are intertwined in both ground and excited states, and they might share the same momentum- and energy spaces. The recent EELS experiments, however, have not been able to detect well-defined CDW excitations[104,105], while acoustic plasmons have been observed by RIXS using both Cu $L_3$ edge and O $K$ edge[106,107]. The observed acoustic plasmons, characteristic of layered metals with weak interlayer coupling, disperse steeply to the energy of ~ 1.0 eV, and hence confined in the small $q$ region ($q < 0.1$ r.l.u.). They are outside the region of the expected collective CDW excitations, typically for $q = 0.2 – 0.3$, so that both charge modes do not interfere.

The recent RIXS studies of Bi2212 and Bi2201 have suggested the presence of dispersive CDW excitations emanating from $Q_{CDW}$ that intersect with optical phonon branches near $Q_{CDW}$, thereby generating the phonon anomalies discussed below. Likewise, for the stripe-ordered LBCO, an ultrafast spectroscopy, time-resolved resonant x-ray scattering, has detected the presence of gapless excitations, possibly corresponding to the transverse stripe fluctuations (meandering mode) and slow dynamics of topological defects.

Relatively high-energy charge excitations have also been observed in RIXS on Hg1201[66] and electron doped cuprates[94] that extend to 0.2 – 0.3 eV, although it is not clear whether they are collective or electron-hole excitation continuum. For Hg1201 an additional excitation peak has been identified in RIXS, but no corresponding feature has been seen for other cuprates, *e.g.*, YBCO.

So far, the spectroscopies have been unable to detect clear signals from the CDW collective modes. It may be that these two spectroscopies do not have sufficient $q$- and $\omega$- resolution at the present stage. Alternatively, the CDW excitations may be strongly damped and not well-defined, or they may be difficult to observe as pure charge excitations due to intertwining with spin and/or pairing correlations.

*CDW gap*

These difficulties exist also in the observation of the CDW gap using RIXS and EELS. Not only the modern RIXS and EELS but also optical spectroscopy have been unsuccessful in observing a gap unambiguously associated with the CDW order even for the stripe ordered La214 which shows clear features in the $T$ dependence of the in-plane and $c$-axis resistivity at the onsets of the charge- and spin-stripe orders as already seen in Fig. 2(a). Nevertheless, a fairly large gap is expected from the very short CDW correlation length of about one CDW period as in the two reference systems, BaBiO$_3$ and 14-24-41 cuprate. In the bismuthate an extremely large CDW gap of 2 eV is clearly observed, so that the correlation length is almost the distance of the nearest neighbor Bi atoms[76] - an extremely short correlation length $\xi$ of CDW is deduced from the observed large CDW gap ($2\Delta_{CDW}$) ($\xi = \hbar v_F/\Delta_{CDW}$, $v_F$ being the Fermi velocity) which corresponds to a mean-field CDW ordering temperature $T_{CDW}$ ~ 2500 K. One may expect strong CDW fluctuations in this system, but the static CDW order is quite stable, persisting to, at lowest, 1000 K due to three-dimensionality and the robust breathing mode distortions of the oxygen atoms. The charge order in the 14-24-41 cuprate is characterized with the gap of ~ 0.1 eV[78], nearly the same size as the pseudogap in 2D cuprates. Like the 2D-CDW in non-La214 cuprates, the onset of the charge order in the ladder cuprate is gradual and at temperature, typically ~ 100 K, much lower than the mean-field value expected from the gap size reflecting its low dimensionality.

From the early stage of high-$T_c$ research, the traditional spectroscopies, optical, Raman, photoemission, and STS, have detected gaps in the one-particle and two-particle (electron-hole) excitations. The observed gaps have been assigned either to $d$-SC gaps or to the normal-state pseudogap. Recent Raman scattering measurements on YBCO and Hg-based cuprates have reported a possible observation of a CDW gap, and found the energy scale of CDW gap excitations similar to that of d-wave SC gap at antinodes and even comparable to the PG energy scale (~ 0.1 eV)[108]. Considering that we are dealing with intertwined orders, the CDW gap may not exist in a pure form. It is possible that the so far observed gap is a spin-charge-pair entangled gap. An example is a gap observed for the hole-doped two-leg ladder cuprate. In this cuprate a gap in the optical spectrum develops associated with a gradual onset of charge order. This gap is not a simple charge order gap. It shows up as a dissociation gap of hole pairs, and also as a spin gap in the spin excitations.

**3.3 Phonon anomalies**

The so-called phonon anomalies refer to softening and/or linewidth broadening at or near $Q_{CDW}$ in the Cu-O bond direction, analogous to the Kohn anomaly in classical CDW systems which arises from strong electron-phonon coupling (EPC). Previous studies focused on Cu-O bond stretching (BS) (and bond buckling phonon modes) which are expected to relatively strongly couple to the underlying electronic states, and are presumably responsible for the kinks in the electron dispersions observed by the angle-resolved-photoemission-spectroscopy (ARPES)[109,110]. In fact, the phonon softening in the BS branch is ubiquitously observed for most of the cuprate families so far studied. However, this anomaly in cuprates differs from the Kohn anomaly in that the BS phonon softening occurs in a broad range of momentum around $Q_{CDW}$ and that the phonon energy decreases at largest 20 - 30 % from the original one (typically ~ 60 - 70 meV), never reaching zero energy even at $T_{CDW}$ and at lower temperatures. Obviously, the anomaly is associated with CDW, and there exists sizable EPC.

There have been several reports that observed similar phonon anomalies also in other branches, including the low-energy acoustic and optical phonon dispersions. The detailed crystallographic refinements of the CDW state in YBCO show significant displacements of most of the atoms in the unit cell[111], so that it is possible that many phonon branches show similar anomalies. However, the anomalies in the phonon branches lower in energies than the BS phonon are not necessarily related to CDW order or CDW fluctuations. They are also explicable as a hybridization with other phonon branches close in energy and crossing each other at $q$ near $Q_{CDW}$ (branch anti-crossing)[112].

A question is then how phonons are modified in the presence of CDW. Important hints have been provided by the recent experimental studies[65]. First, the BS phonon softening is already apparent at temperature far above $T_{CDW}$. Secondly, upon cooling below $T_c$, whereas the 2D-CDW is weakened, the softening is more pronounced. These findings strongly suggest that the CDW (or stripe) fluctuations or dynamic CDW correlations, rather than the static or quasi-static charge order, are responsible for the phonon anomaly, arising from a strong *dynamical* coupling of the BS phonon with dynamic CDW fluctuations and resulting in a significant renormalization of the phonon energy and lifetime. This is in line with an idea that the phonon softening is generated by a coupling to collective CDW excitations which steeply disperse out from $Q_{CDW}$ and intersect the BS phonon branches near $Q_{CDW}$[71,113]. The same scenario of dynamical coupling between optical phonons and fluctuating stripe was theoretically proposed early on for a stripe system[92]. Thus, the phonon anomalies are likely view as a manifestation of dynamic CDW fluctuations.

*Doping dependence of the BS phonon softening*

Given that the BS phonon softening are imprints of the CDW fluctuations on the phonon dispersions, their doping dependence would provide information about how the strength of CDW fluctuations vary with doping. For the cuprates so far investigated, the BS phonon softening is a robust feature, persisting to temperature far above $T_{CDW}$, in the doping range $0.12 < p < 0.2$, and strongest near optimal doping[114]. The doping dependence is distinct from that of the ARPES-kink which is appreciable even for undoped and heavily overdoped cuprates, pointing to the different origin.

The result indicates that the CDW fluctuations may be strongest at doping where $T_c$ is highest, while the quasi-static CDW order is strongest at lower doping $p \sim 1/8$, suggestive of a constructitive relationship between SC and CDW fluctuations. However, this must be checked by a more quantitative RIXS investigation of the doping evolution of the charge density fluctuations.

### 3.4 Relation to spin fluctuations

It is well known that the spin excitation spectrum or the dispersion of spin excitations is similar in all cuprate families, either La214 or non-La214[115]. The undoped cuprates with commensurate AF order show spin wave (magnon) dispersion from (0.5, 0.5) in r.l.u. up to a high energy of ~ 0.3 eV reflecting extremely large superexchange coupling $J$ (100 - 150 meV). With hole doping, the magnons evolve into heavily damped excitations (called 'paramagnons') dispersing upward from a characteristic energy $\omega_r$. Below $\omega_r$, the excitations are less damped with downward dispersion toward incommensurate spin ordering wavevectors $(0.5 \pm \delta_s)$, $\delta_s = Q_{SDW}$. This "hourglass" shaped dispersion of spin excitations are commonly observed in cuprates, regardless of La214 and non-214. A difference between 214 and non-214 is the presence of a spin gap below $\omega_r$ in the latter. Underdoped Hg1201 is seemingly an exception in which the dispersion looks like a "wine glass" without no clear splitting of the dispersion below $\omega_r$ [116]. In view of the dispersion that restores a hourglass shape in a more hole doped sample below $T_c$ [117], the low-energy spin excitations in Hg1201 are subject to stronger damping than other cuprates, another peculiarity of Hg1201.

The low-energy inelastic spin excitations are gapless in La214 showing stripe order and in LSCO for $x < 0.14$. The spin (stripe or SDW) incommensurability $\delta_s$ of the gapless excitations increases linearly with $x (= p)$[17,18], as expected from the strongly correlated stripe picture. The lowest-energy spin excitations associated with the SDW order in heavily underdoped YBCO ($p < 0.09$) shows a similar trend, $\delta_s \sim p$ [118,119]. As $p$ exceeds 0.09, a gap in the spin excitations (spin gap) and simultaneously CDW correlations without any related SDW correlations start to develop. This does not imply that the spin fluctuations play a secondary role in this doping region. The short-range spin interaction is little affected by increasing doping[120], and the antiferromagnetic (AF) correlations are still intact up to slightly overdoped region[121]. Since the dynamic spin fluctuations in cuprates are much stronger than in conventional metals, there is a school of thought that the charge fluctuations play a secondary role and that the charge order is a consequence of strong AF spin correlations[122,123].

In the phase diagram of non-214 cuprates, both CDW and SDW (or AF) orders coexist with the SC order. A question is whether fluctuations of both orders drive the superconductivity, rather than spin fluctuations alone. Note that the CDW order does not share the same doping region with the SDW order (spin-charge decoupling). Then, similar spin-charge decoupling is expected in the fluctuation spectra not only for non-214, but also for the stripe-ordered La214 in view of the tendency for spin-charge decoupling becomes apparent with elevating temperature[63]). The spectrum of charge fluctuations in the $\omega$-$q$ plane, the exact shape of which is vague at present, extends to the energy scale of 0.2 – 0.3 eV, comparable to that of the hourglass-shaped spin fluctuations. To further speculate, if the spin fluctuations contributed to the pair formation as a glue, charge fluctuations would play either as an independent glue to enhance SC order[124,125] or as a pair breaker for $d$-wave pairing.

### 3.5 Relevance to other phases

*Pseudogap and pair-density-wave*

The origin of the PG continues to be a matter of hot debate. There have been circumstantial evidences that suggest a secondary role of CDW in the PG regime; i) in the phase diagram the CDW order exists in the doping region narrower than that of PG and $T_{CDW}$ is always lower than $T^*$, and ii) the CDW order is weak accompanied by very small lattice deformation[111] and no discernible change at $T_{CDW}$ in charge transport (Fig. 2(b)). The recent experimental findings challenge this view. The universally observed CDW fluctuations and their high onset temperature and high energy scale are the features in common with PG. However, CDW alone cannot explain the vast PG phenomenology which requires a necessity for collaboration or intertwining with other orders. The generic phase diagram also indicates that the PG regime is a playground for short-range and fluctuating charge, spin, and pairing order that, in principle, can create a gap. At present, PDW is only a known candidate for such an intertwined order[31]. We have seen that the 2D superconductivity in the stripe ordered cuprates is best explained by the presence of the PDW in the $CuO_2$ plane. For non-La214 cuprates, a series of STS experiments on Bi2212 have yielded evidences for spatial modulation of superfluid density and/or superconducting gap magnitude superposed on the uniform superfluid (and/or uniform superconducting gap) even at zero magnetic field[126,127]. As described in 2.4, a robust SC correlation remaining in the magnetic-field-induced 3D-CDW phase[56,86,87] is also discussed in the context of PDW.

In addition, an ARPES measurement on Bi2201 observed a new gap feature opening near the antinode[128], usually assigned to the pseudogap. Distinct from the superconducting gap, the gap opens at momentum $k$ distant from the Fermi momentum $k_F$ of the normal metallic state above the PG temperature $T^*$. It was found that the gap closes from below on going from antinodes to nodes which is not explained by a CDW gap, but consistent with a PDW gap[129,130].

*Nematic phase*

There are several experiments that show evidence for a thermodynamic phase transition to a nematic ordered phase with spontaneous breaking of rotational symmetry at $T^*$[131]. A close link of the electronic nematicity with PG is shown by the microscopic STS imaging in Bi2212[132], and a nematic phase transition is suggested also by the emergent anisotropy in the charge transport in YBCO[133]. Nevertheless, the most serious problem with the nematic order is that a nematic order is an intra-unit-cell ($q = 0$) order, which alone cannot open a gap. As a compromise, the nematic order is supposed to be an extremely short-range order of a uniaxial CDW, called a vestigial nematic order[43,134]. However, there remains a few difficulties in order to relate the short-range unidirectional CDW order to the nematic order[135] including the Cu-O bond diagonal nematic order observed for Hg1201[136].

Nematicity is not a phenomenon only in the charge sector. In the lower doping region of the PG antiferromagnetic (AF) correlation prevails, and a spontaneous alignment of the direction of the SDW wavevector $Q_{SDW}$ was observed below about 150 K by the neutron scattering study of YBCO[137]. Since a spontaneous alignment of the short-range 2D-CDW domains does not occur in YBCO, the anisotropy in charge transport is likely linked to the spin nematicity.

*Peculiarities in Hg1201*

Hg1201 has a particularly simple tetragonal structure and the highest $T_c$ among the single-layer cuprates. However, Hg1201 shows several peculiar features in the charge ordering and related phenomenon: i) The correlation length $\xi$ is extremely short over the entire temperature range. Even below $T_{CDW}$ a tendency of growing CDW correlation is not seen. ii) The CDW strength is scarecely affected by the onset of SC, iii) The spin fluctuation spectrum, in particular its lower energy part, is anomalously damped, so that it looks like a wine-glass rather than the hourglass-shaped spectrum universally observed in other cuprates. iv) The direction of the nematicity (a nematic order identified below the PG onset $T^*$) is diagonal, distinct from the nematicity in the Cu-O bond direction in most other cuprates. The reasons for these peculiarities are not known at present. At least, the two characteristic features of Hg-based cuprtates might be related to the peculiarities. First, the energy scales of both SC and CDW in Hg1201 are exclusively large. The SC energy gap scale (and certainly largest SC phase stiffness $\rho_s$) is largest and perhaps this is the case with CDW, linked to the intrinsic short correlation length. Secondly, The disorder effect on both SC and CDW is apparently minimal in Hg1201. Together with supposedly weakest effect of underlying lattice, the fluctuating CDW is hard to be pinned also in its direction.

## 5. Summary and Outlook

Based on the experimental observations in the past few years, is summarize how the CDW correlations evolve in the order of descending temperature.

(1) *High-temperature dynamic CDW correlations*

The experiments performed in the last three years have revealed the almost universal presence of dynamic CDW correlations (PCDW) across different cuprate families with high onset temperature and high energy scale comparable to those of the PG and spin fluctuations. Given that the PG and spin fluctuations are most likely caused by the correlations local in real space, this provides support a strong coupling scenario that the cuprate charge ordering is driven by strong electronic correlations. The high-temperature correlations are essentially two-dimensional and exist not only in a large region of the $T$-$p$ phase diagram. They also involve a large manifold momentum centered near $Q_{CDW}$. Furthermore, they seem to be allowed to dynamically point to any direction, making them to have much larger spectral weight than that of quasistatic 2D-CDW. PCDW is thus characterized by dynamical nearest neighbor charge correlations in real space with correlation length of one to two lattice constant (they are dynamical analog to the static CDW correlations in $BaBiO_3$).

(2) *Low-temperature 2D CDW and 3D stripe order*

Diversity of charge ordering shows up with lowering temperature. Upon cooling, quasistatic 2D short-range CDW order gradually develops below $T_{CDW}$ in non-La214 cuprates. On the other hand, in La214 cuprates with robust stripe order, a static 3D long-range order is established upon the onset of the spin-stripe order. In both cases the formation of 2D-CDW or 3D stripe order is triggered by a lock-in of charge density modulation at a specific $|q| = Q_{CDW}$ along the specific Cu-O bond directions. Consequently, only a small fraction of the spectral weight of the dynamical correlations condenses into the low-temperature order. The lock-in of $Q_{CDW}$ takes place by pinning the fluctuating CDW/stripe by the LTT structure and coexisting spin-stripe order in the case of stripe ordered La214 and possibly by lattice imperfections and material-specific electronic structure in non-La214 cuprates.

(3) *CDW order in the presence of superconductivity*

It becomes progressively clear that any type of CDW order coexists with the SC order (or pairing correlation). *The dynamic CDW correlations or PCDW do not show any discernible change below the SC*. On the other hand, the relationship between the low-temperature CDW order and SC is material dependent. In a class of the cuprate families, *e.g.*, YBCO and Bi2212, the 2D-CDW rapidly weakens below the onset temperature $T_c$ of superconducting order, but the 2D-CDW is not completely suppressed even at temperatures well below $T_c$, coexisting with SC. In another class which includes Hg1201 and electron-doped cuprates, the 2D-CDW and SC orders appear to coexist without any appreciable influence on each other. The relationship between SC and 3D stripe order in LBCO is quite intriguing. While 3D-SC order or the interlayer phase coherence is almost completely suppressed by the 3D stripe order, 2D-SC within each $CuO_2$ plane persists, implying that the long range 3D stripe order coexists with 2D-SC, possibly in the form of PDW.

3D-CDW in YBCO, appearing once superconductivity weakens with magnetic field or uniaxial strain, exhibits a unique coexistence with SC which is as yet fully understood. Despite the apparent competition with SC, as in the case of thermodynamic competition between two 3D long-range ordered phases[56,83], *the two mutually exclusive orders, 3D-CDW and 3D-SC, seem to establish some form of cooperation in order to coexist at low temperature*[56]. In the case of the magnetic-field-induced 3D-CDW this exotic state likely exhibits QOs and hence FRS. The true nature of this state remains to be elucidated, but a PDW state is proposed as a possible candidate[86,87].

The observed temperature-evolution of the charge order correlations have resolved most of the questions raised before 2015, as listed in Introduction:

(i) The dynamic correlations observed at high temperatures are universal, irrespective of CDW in non-La214 or stripes in La214. The distinction between the two becomes apparent at low temperatures due to material specific properties.

(ii) The charge density modulations of the 2D-CDW are unidirectional like the charge stripe order in La214.

(iii) The spin-charge stripe order, the short-range 2D-CDW order, and the magnetic-field-(and uniaxial strain-) induced 3D-CDW are all static on the time scale longer than $10^{-9}$ second (a time scale of the NMR measurement).

(iv) The universal CDW correlations with high-temperature and high-energy scales as well as an intimate relationship with PG point toward the strong electronic correlations as a primary driving force of the cuprate charge order.

(v) The question on the role of the charge correlations or the charge order in the cuprate superconductivity or the formation of a pseudogap remains open.

It is widely agreed on that the charge correlations alone cannot explain the high-$T_c$ superconductivity, and that the charge order is likely a consequence of a hidden primary order such as PDW that is responsible for PG. However, given that both SC and PG in the cuprates might emerge from the intertwining of various types of orders related to spin, charge, pair, and possibly lattice, the charge order and its fluctuations are now recogenized as an important and probably indispensable ingredient in the cuprates physics.

These remarkable advances of our understanding of the cuprate charge ordering owed largely to the progress in the resonant x-ray scattering spectroscpies, RXS and RIXS as well as in the theoretical calculations of the charge order and its interplay with SC in the models of strongly correlated electrons. Then, are the charge orders and their fluctuations are a remaining piece to resolve the high-$T_c$ puzzle? Unfortunately, there still remains a distance toward the final goal. The enigmatic PG is not understood by the charge order correlations alone, but they likely contribute by being intertwined with spin and pairing correlations.

For a full understanding of the charge fluctuations and their role in high-$T_c$ superconductivity, further development of the new spectroscopies, RIXS and momentum-resolved EELS (MEELS), with much improved momentum and energy resolutions are indispensable. They will make it possible to unravel the dynamical charge structure factor over wide momentum and energy spaces. The dynamic structure factor is directly related to the imaginary part of dynamic charge susceptibility associated with collective charge fluctuations that potentially uncover detailed information about the interactions that drive the CDW instability and possibly contribution of charge fluctuations to superconducting pairing. These spectroscopies are expected to play a role, similar to what ARPES and STS did previously, in advancing unified understanding of high-$T_c$ superconductivity in the cuprates and other strongly correlated superconducting systems including Fe-bases superconductors.